# Identification of carbon dioxide in an exoplanet atmosphere

JWST Transiting Exoplanet Community Early Release Science Team[*]

**Carbon dioxide ($CO_2$) is a key chemical species that is found in a wide range of planetary atmospheres. In the context of exoplanets, $CO_2$ is an indicator of the metal enrichment (i.e., elements heavier than helium, also called "metallicity")[1-3], and thus formation processes of the primary atmospheres of hot gas giants[4-6]. It is also one of the most promising species to detect in the secondary atmospheres of terrestrial exoplanets[7-9]. Previous photometric measurements of transiting planets with the Spitzer Space Telescope have given hints of the presence of $CO_2$ but have not yielded definitive detections due to the lack of unambiguous spectroscopic identification[10-12]. Here we present the detection of $CO_2$ in the atmosphere of the gas giant exoplanet WASP-39b from transmission spectroscopy observations obtained with JWST as part of the Early Release Science Program (ERS)[13,14]. The data used in this study span 3.0 - 5.5 µm in wavelength and show a prominent $CO_2$ absorption feature at 4.3 µm (26σ significance). The overall spectrum is well matched by one-dimensional, 10x solar metallicity models that assume radiative-convective-thermochemical equilibrium and have moderate cloud opacity. These models predict that the atmosphere should have water, carbon monoxide, and hydrogen sulfide in addition to $CO_2$, but little methane. Furthermore, we also tentatively detect a small absorption feature near 4.0 µm that is not reproduced by these models.**

WASP-39b is a hot (planetary equilibrium temperature of 1170 K assuming zero albedo and full heat redistribution), transiting exoplanet that orbits a G7-type star with a period of 4.055 days[15]. The planet has approximately the same mass as Saturn ($M = 0.28$ $M_{Jup}$) but is ~50% larger ($R = 1.28$ $R_{Jup}$), likely due to the high level of irradiation it receives from its host star[16-18]. We chose this planet for the JWST ERS transmission spectroscopy observations because analyses of existing space- and ground-based data detected large spectral features and showed that there was minimal contamination of the planetary signal from stellar activity[10,19-21]. The main spectral features previously detected were confidently attributed to sodium, potassium, and water vapor absorption[10,19-20], while $CO_2$ was suggested to explain the deep transit at 4.5 µm seen with Spitzer[10].

Atmospheric metallicity has long been thought to be a diagnostic of the relative accretion of solids and gas during the formation of gas giant planets, both of which bring heavy elements to the hydrogen-dominated envelope and visible atmosphere[4-6]. The metallicity of WASP-39b's host star, which is a proxy for the metal enrichment of the protoplanetary disk that the planet formed in, is approximately solar[15,22-24]. Therefore, the planet mass - atmospheric metallicity trend observed in the solar system giants[25,26] predicts that it has an enhancement of ~10x solar (like that of Saturn, Ref. [27]). Additionally, interior structure models that match WASP-39b's low density predict a 95th percentile upper limit for the atmospheric metallicity of 55x solar, under the limiting assumption that the planet has no heavy element core and that all the metals are evenly distributed throughout the envelope[28].

---
[*] A list of authors and their affiliations appears at the end of the paper.

Despite having some of the highest signal-to-noise detections of spectral features in its transmission spectrum, modeling of the existing data for WASP-39b has resulted in metallicity estimates ranging across five orders of magnitude, from 0.003x to 300x solar[10, 29-33]. The wide range of values stems from the data being of insufficient quality to break the degeneracy between clouds and metallicity in transmission spectra models[34], as well as uncertainty over the interpretation of the photometric measurements by the Spitzer Space Telescope at 3.6 and 4.5 µm. Thus, spectroscopic data with greater precision, finer spectral channels, and wider wavelength coverage were needed to better constrain the metallicity of this (and other) giant exoplanet atmospheres. .

The first JWST ERS observation of WASP-39b was obtained using the Near Infrared Spectrograph (NIRSpec)[35,36] on July 10, 2022, between 15:24 and 23:37 UTC. We used the Bright Object Time Series (BOTS) mode with the 1.6" x 1.6" fixed slit aperture and the PRISM disperser to capture spectra between 0.5 and 5.5 µm. The data were recorded using the SUB512 subarray with five groups per integration and the NRSRAPID readout pattern, which gave integration times of 1.38 s. NIRSpec obtained a total of 21,500 integrations over 8.23 hours of observations centered on the 2.8 hour transit duration of WASP-39b.

The count rate in the PRISM mode varies significantly over the bandpass due to the spectral energy distribution of the star and the wavelength-dependency of the spectrograph dispersion. Therefore, the observations were designed to saturate at shorter wavelengths in order to obtain sufficient signal-to-noise ratio at the longer wavelengths in the bandpass for first-of-its kind spectroscopy of transiting exoplanets. Wavelengths between 0.71 and 2.09 µm have at least one group saturated in the pixel at the center of the spectral trace. We concentrate here on the analysis of the data longward of 3.0 µm that are not impacted by saturation to investigate the spectrum overlapping with the previous 3.6 µm and 4.5 µm Spitzer photometric measurements. The subset of the PRISM data described herein has a native spectral resolving power ($R=\lambda/\Delta\lambda$, where $\lambda$ is wavelength) of 100 - 350. For this study, we binned the data to lower resolving powers (values range from 60 to 200 depending on wavelength and reduction). The binning is done at the light curve level before the fitting of the transit depths that constitute the transmission spectrum. Analyses of JWST/NIRSpec transit observations obtained during commissioning have shown that similar levels of binning as we use here results in minimal systematics[37]. An analysis of the complete PRISM dataset at full resolution including recovery of the saturated part of the spectrum is ongoing.

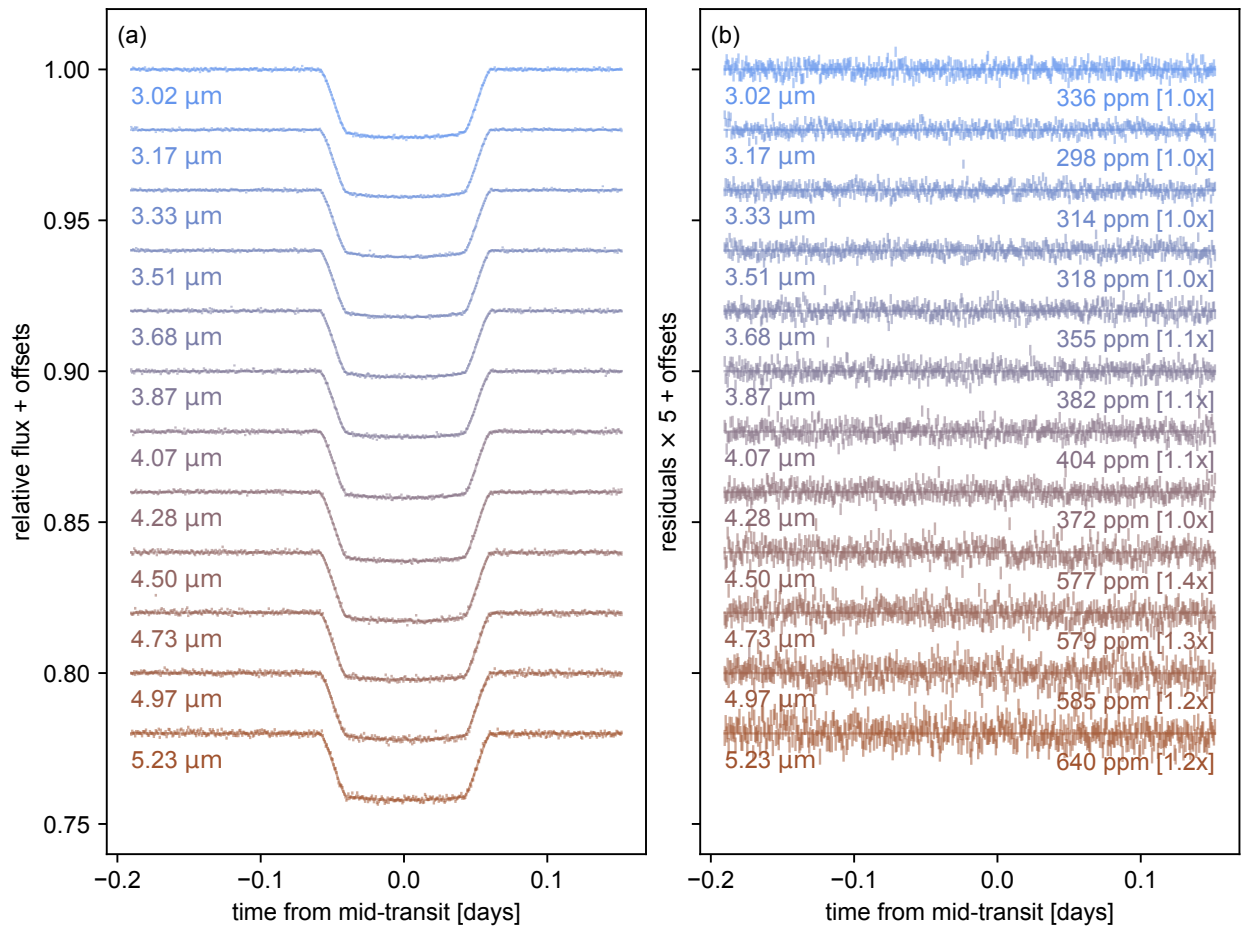

**Figure 1: JWST NIRSpec time-series data for WASP-39b.** a) Spectroscopic light curves for WASP-39b's transit with a spectral resolving power of 20 and a time cadence of 1 minute (data are binned and offset vertically for display purposes only). An exoplanet light curve model was fitted to the data using a quadratic limb darkening law with an exponential ramp and a quadratic function of time removed. b) Residuals of the binned light curve after subtracting the transit model scaled up by a factor of five to show the structure. The RMS of the residuals are given in units of ppm. The numbers in brackets are the ratio of the RMS to the predicted photon-limited noise.

We reduced the NIRSpec PRISM data for WASP-39b using the JWST Science Calibration Pipeline along with customized routines to minimize noise in the time series spectra (see Methods). We performed four different reductions of the transmission spectrum starting from the uncalibrated data[21, 38-40]. Figure 1 shows derived spectroscopic transit light curves from one of the reductions.

We confirm with our analysis of the WASP-39b data that NIRSpec transit observations at a resolving power of 60 - 200 are nearly free of systematics. We achieved close to photon-noise-limited measurements in the spectroscopic light curves after trimming the first 10 minutes of data and removing a linear trend in time with an average rate of ~190 ppm/hour across the bandpass. We also obtained similar results by fitting the full time-series with a downward trending exponential ramp (timescale ~100 minutes) combined with a quadratic function of time. The lack of large systematics in these data stands in contrast to previous transit spectroscopy observations with space- or ground-based telescopes[41].

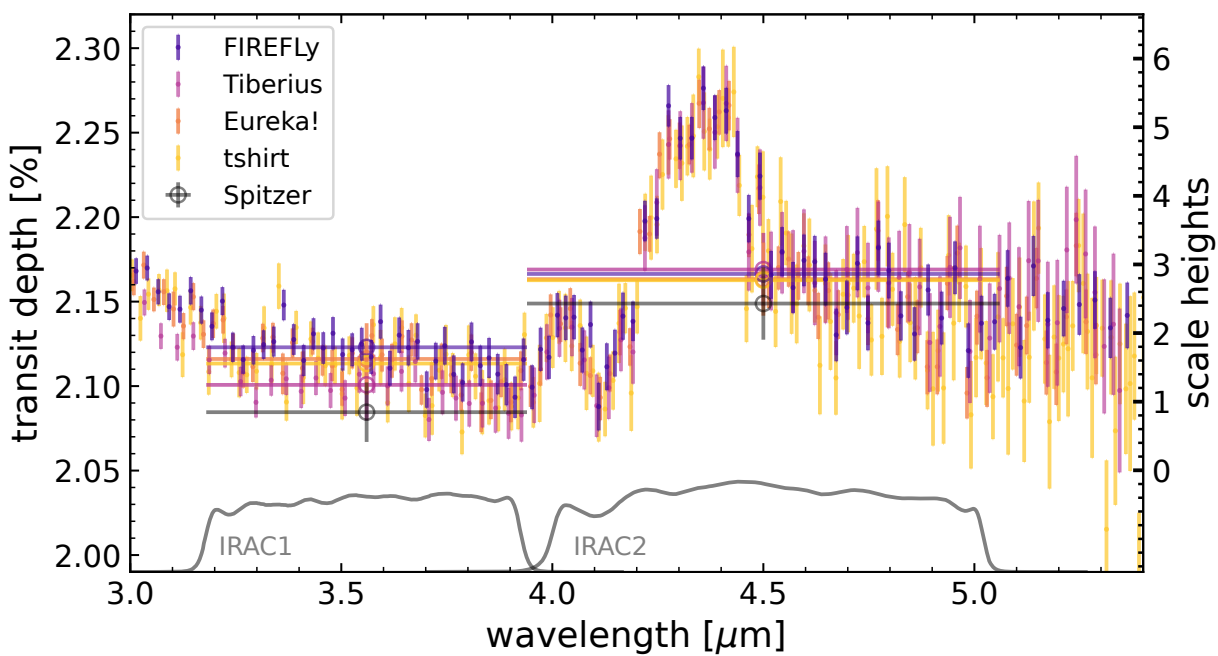

**Figure 2: Independent reductions of the WASP-39b transmission spectrum.** The JWST data (small colored points) are compared to Spitzer's two broadband photometric measurements (grey circles and corresponding sensitivity curves labeled "IRAC1" and "IRAC2"). The axis on the right shows equivalent scale heights (750 - 1000 km) in WASP-39b's atmosphere; for plotting purposes we assume that one scale height corresponds to 800 km. The JWST data are consistent with the Spitzer points (within 2σ) when integrated over the broad bandpasses (indicated by the horizontal lines). The relative transit depths between the 3.6 and 4.5 µm channels are also consistent within 2σ between independent reductions of the JWST data, with most of the deviation coming from the 3.6 µm bandpass. Vertical error bars indicate 1σ uncertainties.

The transmission spectra derived from the different reductions, shown in Figure 2, have excellent agreement. They all exhibit a large feature at 4.3 µm, as well as a smaller feature near 4.0 µm (discussed below). Detailed modeling of the FIREFLy-reduced data yields a statistical significance of 26σ for the large feature (see the Methods). We attribute this feature to $CO_2$ absorption based on a comparison of the resolved band shape to theoretical models and the spectra of brown dwarfs[42]. Figure 2 also includes Spitzer's two broadband photometric measurements[10], which are consistent with the JWST data to better than 2σ after integrating the transmission spectrum over the Spitzer bandpasses. We also see good agreement (better than 2σ for all reductions) in the relative transit depths between the 3.6 and 4.5 µm channels. The comparison shown in Figure 2 demonstrates both the consistency in the derived spectra from multiple, independent analyses and the reliability of the previous Spitzer measurements.

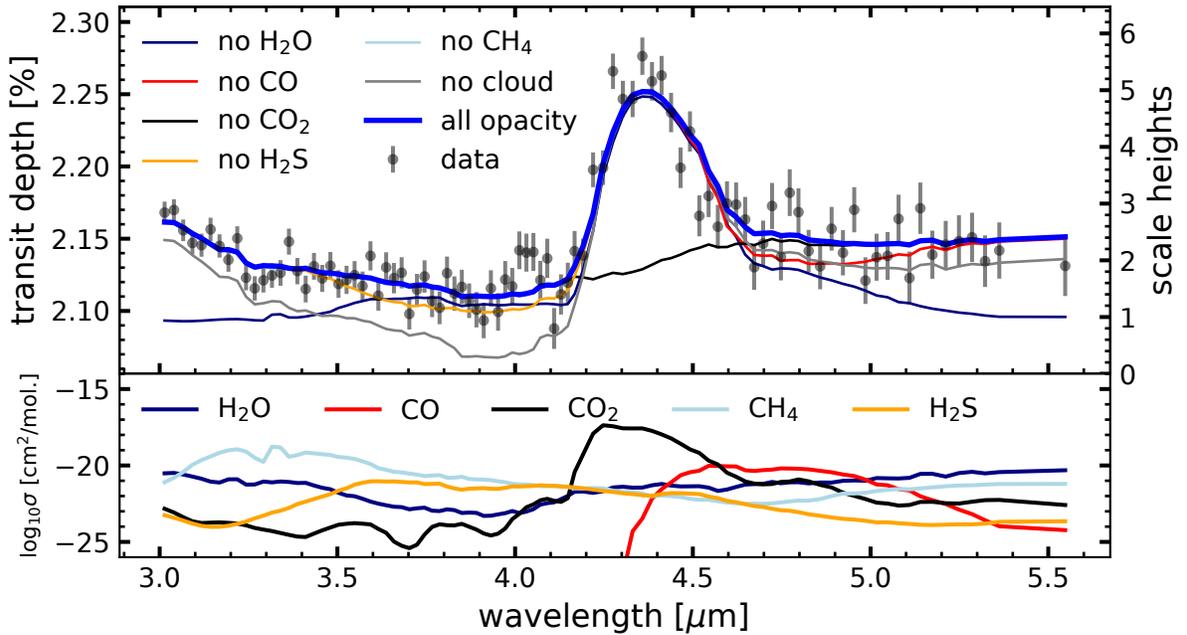

**Figure 3: Interpretation of WASP-39b's transmission spectrum.** The top panel shows a comparison of the FIREFLy reduction to the best-fit ScCHIMERA theoretical model binned to the resolution of the data (blue curve, see Methods). The key parameters of the model are 10x solar metallicity, carbon-to-oxygen ratio of 0.35, and cloud opacity of $7 \times 10^{-3}$ cm$^2$/g. The impact of the opacity sources expected from thermochemical equilibrium over the full bandpass are indicated by removing the opacity contribution from individual gasses one at a time. As in Figure 2, the axis on the right shows equivalent scale heights in WASP-39b's atmosphere. The bottom panel shows the molecular absorption cross-sections for each gas in the best-fit model. The model is well matched to the data ($\chi^2/N_{data}$=1.3), suggesting that our assumptions broadly capture the important physics and chemistry in WASP-39b's atmosphere. However, there is a feature near 4.0 µm that cannot be reproduced by the models used here. The strong $CO_2$ absorption (4.1 - 4.6 µm) and the apparent lack of methane (3.0 - 3.5 µm) is what drives the solution to an elevated atmospheric metal enrichment, ruling out previous low metallicity estimates[29-31]. The other reductions and models give similar results.

We compared the data to a suite of one-dimensional atmospheric structure and transmission spectrum models to constrain the composition of WASP-39b's atmosphere. These models assume radiative-convective-thermochemical equilibrium, and they adopt a scaled solar abundance pattern. We calculated planet-specific grids of these models over a range of atmospheric metallicities, carbon-to-oxygen ratios, and cloud properties using four different codes. These grids of self-consistent model transmission spectra were then fitted to the FIREFLy-reduced data (the fit results are independent of which data set we use) while also adjusting for a reference radius at 1 bar. The results are illustrated in Figure 3; see the Methods for further details.

Under similar assumptions, all four model grids are able to match the dominant spectral morphologies - namely the strong $CO_2$ feature between 4.1 and 4.6 µm and the rise in transit depth blueward of 3.6 µm due to water vapor (a species that had been detected previously at shorter wavelengths, Ref. [10]). More subtle modulations over the whole bandpass are potentially due to contributions from clouds, carbon monoxide, and hydrogen sulfide, though the degree to which the two gas species contribute is unknown pending further study.

Several models for warm gas giant atmospheres predict that the $CO_2$ abundance scales quadratically with atmospheric metallicity, becoming detectable at 4.3 µm for metallicities above that of the Sun[1-3]. The representative best-fit model shown in Figure 3 is consistent with this scenario. It has a 10x solar metal enrichment and a slightly sub-solar carbon-to-oxygen ratio (0.35, compared to the solar value of 0.55; Ref. [43]). The moderate contribution of cloud opacity predicted by the best-fit model is consistent with interpretations of previous population-level studies of planets that have similar temperatures and gravities as WASP-39b[44,45]. It is also consistent with the predictions of aerosol microphysics and global circulation models of hot giant planets[46-48].

In addition to the large $CO_2$ feature, we also identify a smaller spectral feature near 4.0 µm that is not matched by our thermochemical equilibrium models (see Figure 3). This feature is present in all four independent reductions and has a significance of 2σ (see the Methods). Further data analysis and modeling including non-equilibrium chemistry are needed to fully assess the robustness of this feature and to identify the chemical species that gives rise to it. Additional JWST ERS observations of WASP-39b that will use the G395H grating on NIRSpec also have the potential to confirm the 4.0 µm feature and resolve it in greater detail.

The grid fits explored here favor lower metallicities than Refs. [10, 21], and higher metallicities than Ref [31], even though the Spitzer data that their studies included are consistent with our JWST data. The higher precision and more resolved measurement of the $CO_2$ feature enabled by JWST pulls the models of Refs. [10, 21] to lower metallicity and increased cloudiness. Nevertheless, it is not possible to obtain a robust confidence interval on this inference without more rigorous Bayesian analyses, which is left to future work (see Methods). Continued modeling of WASP-39b will also be aided by the future measurements of the planet's transmission spectrum from 0.5 - 5.5 µm that are also being obtained by this Early Release Science program. The final transmission spectrum will ultimately have higher spectral resolution than the data presented here (>4x over most of the bandpass) and will be validated using multiple JWST instruments.

## Methods

**Data reduction**

We reduced the JWST NIRSpec PRISM data for WASP-39b using four separate pipelines to confirm that the results did not depend on the specifics of the analyses, as was sometimes the case for results from the Spitzer Space Telescope (e.g., Ref [49]). The descriptions below refer to calibration pipelines and other software whose code and citations appear in the Code Availability section, below.

**tshirt pipeline**

We used the Time Series Helper and Integration Reduction Tool[40] (tshirt) to extract light curves of the spectrum. This pipeline modifies the JWST Calibration pipeline steps to improve the precision of the reduction. tshirt has been used to successfully analyze the JWST transit observations of HAT-P-14b that were obtained during commissioning with NIRCam[37]. First, we used an updated bias frame from commissioning program 1130 observation 29 and ran the JWST Calibration pipeline until the reference pixels step. We then applied a correction for 1/f-noise which varies for odd and even rows and for each column. We use background pixels for the calibration since reference pixels are not available in this subarray. We skipped the jump and dark subtraction steps because they were seen to add noise to the light curves. tshirt fits the profile of the spectrum with splines and rejects outlier pixels that are more than 50 sigma from the spline fits. We used covariance-weighted extraction[50] with an assumed pixel correlation of 0.08. For spectral extraction, we used a background region no closer than 7 pixels on either side of the source and an extraction region width of 16 pixels. The scatter in the light curve was consistent with the theoretical limit of photon and read noise over short timescales.

We fit the light curves with a second-order (quadratic) polynomial baseline, uninformative quadratic limb darkening priors, and an exponential startup ramp with 10σ clipping of outliers. To begin, we fit the white light curve with priors on the transit center, inclination, and period from Ref [22]. We also used the $a/R_*$ from Ref [22] but widened the uncertainty on this parameter because the enforced prior resulted in significant residuals. Next, we fit each spectroscopic light curve individually with the orbital parameters fixed at the value from the white light posterior medians. We modeled the light curves using the "exoplanet" code[51] and the pymc3[52] sampler. We evaluated the wavelengths using the JWST Calibration pipeline at pixel row 16 (Y=16) from the world coordinate solution. This uses an instrument model and could not be verified due to a lack of strong stellar absorption features at the NIRSpec resolution. All the other reductions adopted this wavelength calibration. As shown in Figure 1, the standard deviation in the out-of-transit light curve approaches the theoretical limit of photon and read noise at short wavelengths, but is 20% to 40% higher at longer wavelengths, which may be related to uncorrected 1/f-noise.

**Eureka! pipeline**

Eureka![39] is a data reduction and analysis pipeline for time-series observations with JWST or HST. Its modular, multi-stage design provides flexibility and ease of comparison at any step, starting from uncalibrated FITS files and resulting in precise transmission or emission spectra. Eureka! has been used to successfully analyze the JWST transit observations of HAT-P-14b that were obtained during commissioning with NIRCam[37].

We began the data reduction process using the UNCAL files available from the MAST archive. The first stage of the Eureka! pipeline is primarily a wrapper for Stage 1 of the JWST Calibration pipeline, which converts groups to slopes. For this dataset, we skipped the jump detection step as it led to a large fraction of detector pixels being incorrectly flagged as outliers. We did, however, search for and flag outliers at multiple points in subsequent stages. We also manually updated the bad-pixel map to include identified hot pixels on the detector that were not provided in the current (July 2022) full-detector STScI data quality map. As part of Eureka!, we performed a custom background subtraction at the group level prior to Stage 1 ramp fitting to account for 1/f-noise introduced during detector readout. We set the top and bottom six rows of the detector as our background region and flagged pixels deemed outliers at $>3\sigma$. We then subtracted the mean flux per pixel column and repeated this for each group and integration in the observation. Similar to Stage 1, the second stage of the Eureka! pipeline is a wrapper for Stage 2 of the JWST Calibration pipeline, which calibrates the 2D time series of fitted slopes. Here, we skipped the flux calibration step, thus leaving the data in units of DN/s.

For Stage 3, we performed background subtraction and optimal extraction of the stellar spectrum for each integration with Eureka!. We only used pixels 14 to 495 in the dispersion direction of the 512x32-pixel subarray, as NIRSpec's throughput is negligible beyond this range. We also masked pixels that have a non-zero data quality flag to avoid any impact of outlier pixels on the extracted spectra or background subtraction. The position of the source on the detector along the cross-dispersion dimension is located by fitting a Gaussian to the pixel values summed over all detector columns. For each pixel, we examined its flux variation in time and performed a double-iteration, $10\sigma$ outlier rejection test. We then executed a second column-by-column background subtraction, this time at the integration level, using pixels located at least 8 pixels away from the source position to compute the mean background per column. Performing this additional background subtraction reduced the number of outliers in the measured light curves and accounted for the residual background and/or noise introduced during the ramp fitting procedure. As with Stage 1, we exclude $3\sigma$ outliers from our background region. We adopted an aperture half-width of 7 pixels for our optimal spectral extraction step, constructing the profile from the median frame. At the end of this stage, we obtained a time series of 1D spectra.

For the remaining stages, we used multiple pipelines (Eureka![39] and ExoTEP[53-55]) to generate and fit the light curves. We first generated median-normalized light curves at the instrument's native resolution (i.e., from each detector column) using our Stage 3 outputs. We then clipped additional outliers in time for the white and spectroscopic light curves. For this step, we first rejected integrations that were more than $3\sigma$ outliers for the source position in the cross-dispersion direction, the width of the fitted Gaussian to the spatial profile, or the drift in the dispersion direction. Next, we produced a median-filtered version of the light curve and clipped out $3\sigma$ outliers in flux. We jointly fit astrophysical and systematics model parameters to the white and individual spectroscopic light curves. Our astrophysical transit model used the batman package[56] with uniform priors, fitting for the following astrophysical parameters: the two coefficients of a stellar quadratic limb-darkening law, impact parameter, semi-major axis, transit time, and the planet-to-star radius ratio ($a/R_*$) in each of the wavelength channels. While the limb-darkening coefficients and planet-to-star radius ratio were fit independently in each spectroscopic channel, we used the best-fitting value of the planet's impact parameter, semi-major axis, and transit time from a white light curve fit as a fixed value in the wavelength-dependent fits. For the systematics model, we assumed a linear trend in time for each wavelength channel, fitting for both the slope and y-intercept. Last, we fit a single-point scatter to each light curve, which illustrates the level of additional noise required for our joint model to reach a reduced chi-squared of unity. The white light curve residuals have an RMS of 3013 ppm and the spectroscopic light curves above 3 μm have a median RMS of 5779 ppm. Similar to the reduction shown in Figure 1, both pipelines reach near photon noise. The Eureka! and ExoTEP transmission spectra appear nearly identical; therefore, only one (Eureka!) is shown in Figure 2.

**Tiberius pipeline**

We built upon the pipeline developed for the analysis of LRG-BEASTS data[21,57,58] to provide an independent reduction of the data. We began with the outputs of the JWST Calibration Stage 1 pipeline with the jump step correction turned off. We created bad-pixel and cosmic-ray masks by identifying 5σ outliers in running medians operating along pixel rows and along individual pixels in time. Prior to tracing the spectra, we interpolated each column of the detector onto a finer grid, 10x the initial spatial resolution, in order to improve the extraction of flux at the sub-pixel level. We used a 4th-order polynomial to trace the spectra and a 4-pixel-wide aperture. To remove the 1/f-noise, we fit a linear polynomial to 21 background pixels along each column in the cross-dispersion direction. Next, to correct for shifts in the dispersion direction, we cross-correlated each stellar spectrum with the first spectrum of the observation to account for very small (0.003 - 0.005) sub-pixel shifts. Our white light curve spans a wavelength range of 0.518–5.348 μm after masking saturated pixels, and our 147 spectroscopic light curves used 3-pixel-wide bins across this same wavelength range. We masked frames 20751–20765 due to a high gain antenna move that led to increased noise in the light curves.

We fit our light curves with a combination of a quadratically limb-darkened transit model (through batman[56]) with a linear-in-time polynomial. We began by fitting the white light curve to derive the system parameters: inclination, $i$, time of mid-transit, $T_C$, the semi-major axis scaled to the stellar radius, $a/R_*$, and the linear limb darkening coefficient, $u_1$. We placed wide boundaries on the parameter values only to prevent unphysical values. In practice, the parameter values did not get close to the boundaries. We fixed the planet's orbital period to 4.0552941 days and eccentricity to 0 from Ref [22]. We fixed the quadratic coefficient, $u_2$, to theoretical values determined by Exo-TiC-LD[59,60] with 3D stellar models[61], and fit for $u_1$. We used a Levenberg-Marquardt algorithm to fit our light curves, rescaled our photometric uncertainties to give a reduced $\chi^2=1$ for our best-fit model and then re-ran the fits. For the spectroscopic light curves, the system parameters ($i, T_C, a/R_*$) were held fixed to the best-fit values found from the white light curve. The white light curve residuals had an RMS of 2761 ppm and the spectroscopic light curve residuals had a median RMS of 6731 ppm. In both cases, the variance of the residuals scales upon binning as expected for Poisson noise.

**FIREFLy pipeline**

We also reduced the data using the Fast InfraRed Exoplanet Fitting Lyghtcurve (FIREFLy) reduction routines[38]. These routines utilize the JWST Calibration pipeline with custom modifications. This pipeline has been used to successfully analyze the JWST transit observations of HAT-P-14b that were obtained during commissioning with NIRSpec G395[37]. We removed 1/f-noise (see Ref [36]) at the group level, as the 1/f-noise changes from group to group. We also skipped the jump step and instead flagged and removed cosmic rays, bad pixels, hot pixels, and other outliers using median filtering of the data both spatially and in time, flagging pixels using a 5σ outlier threshold algorithm. The time series of 2D spectra were aligned using cross-correlation and interpolation, with the time series spectra exhibiting an RMS jitter of 0.005 pixels in the X-axis direction and 0.0026 pixels in the Y-axis direction. We found a small inverse ramp in the light curves, which settled down after the first 2000 exposures, which we discarded. We fit the light curves with the batman[56] transit model along with a linear baseline and a second-order jitter detrending polynomial of X and Y detector position as described by Ref [38], which are present in the spectrophotometry at the 53+/-2 ppm level in the x-direction and 140+/-3 ppm in the y-direction. We applied a fixed quadratic limb darkening law using the 3D models[61] computed using the methods of Ref [62] from ExoTiC-LD[59,60]. In fitting the 3 to 5.5 μm white light curve, we allowed the semi-major axis in units of stellar radii $a/R*$, inclination $i$, and central transit time $T_0$ to freely vary along with the transit depth and systematics model. We used the Markov chain Monte Carlo sampling routine EMCEE[63] to find the best-fit parameters and measure the posterior distribution. We find the 3-5.5 μm white light curve has a transit depth of 2.1368+/-0.0014% and achieves 808 ppm scatter in the residuals. This is within 6% of the expected noise limit of 758 ppm as calculated by the JWST Calibration pipeline, with the scatter of the residuals decreasing to below 40 ppm upon binning with no detectable red noise. We fit each spectroscopic light curve shown in Figure 2 with the same astrophysical and systematic models as the white light curve, except fixing the system parameters ($a/R*$, $i$, $T_0$). The transmission spectral light-curve residuals for each bin are typically within 5% of pipeline error or better, also with no detectable red noise.

**Data-Model Comparison**

We compared the extracted transmission spectral data to a suite of 1D self-consistent radiative-convective-thermochemical equilibrium model atmospheres (see e.g., Refs [64, 65] for a general description of such models) described below. In short, all models are able to fit the 3-5.5 μm spectra consistently (with $\chi^2/N_{data} < 1.4$) with a 10x solar metal enrichment and varying grey cloud opacity for their single best estimate. Comparisons of the model fits from each grid are shown in ED Figure 1. For additional parameters within the grid (e.g., C/O and heat redistribution), there is some discrepancy between each model grid's single best estimate values. Additional Bayesian analyses are needed to rigorously quantify confidence intervals on atmospheric properties of interest, which is beyond the scope of this work. Future works will focus on modeling that includes the effects of disequilibrium chemistry, aerosol microphysics, and three-dimensional circulation effects. We assumed the following parameters in the modeling: stellar $T_{eff}$ = 5512 K, stellar radius = 0.932 $R_{Sun}$, planet mass = 0.281 $M_{Jup}$, planet radius = 1.279 $R_{Jup}$, and planet orbital semi-major axis = 0.04828 AU.

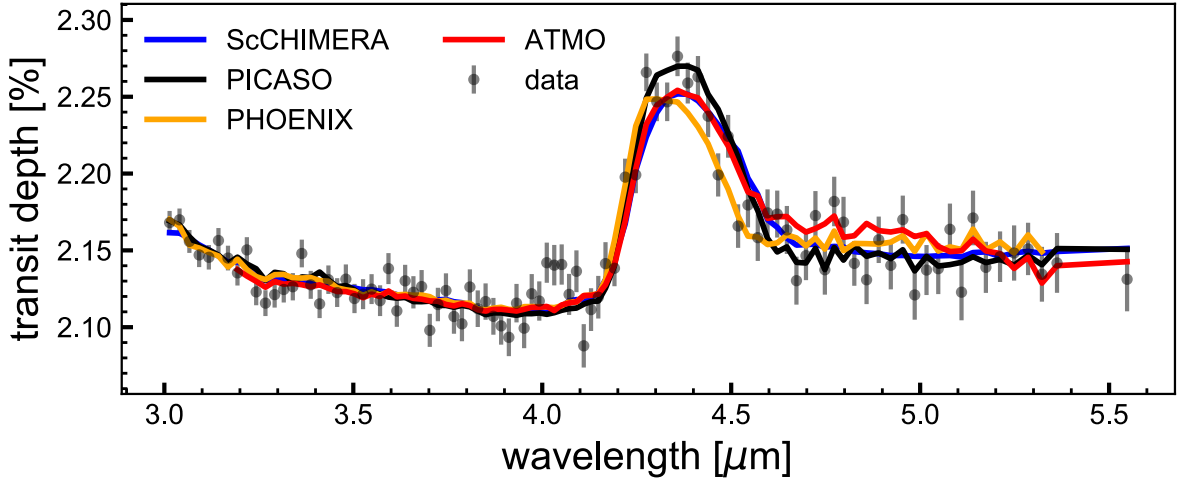

**ED Figure 1: Comparison of transmission spectrum modeling results from different codes for WASP-39b.** Despite different radiative-convective equilibrium and chemical solvers, treatments of clouds, grid spacing, and grid-fitting approaches, all four grids arrive at the same 10x solar metallicity point solution. Additionally, all four provide an acceptable fit to the data, with best fitting $\chi^2/N_{data}< 1.4$.

**ScCHIMERA**

This framework was first described in Refs. [66, 67], with the most recent updates, methods, and opacity sources described in Ref. [68]. We compute the converged atmospheric structure (temperature-pressure and thermochemical equilibrium gas mixing ratio profiles) over a grid of atmospheric metallicity ([M/H], where brackets indicate $\log_{10}$ enrichment relative to solar[43]) spaced at 0.25 dex intervals between 0 and 2.25 (1 to 175 times solar) and carbon-to-oxygen ratio (C/O) at values 0.2, 0.35, 0.55, 0.7, 0.75, 0.8. We assume full day-to-night temperature redistribution[69] as planets in this temperature regime are unlikely to possess strong day-to-night temperature contrast[70,71]. We then compute transmission spectra[72,73] from these converged atmospheric structures. To match the models to the data, the DYNESTY[74] fitting routine is used to search for the optimal [M/H] and C/O (via nearest neighbor) while also simultaneously adjusting the 1 bar planetary radius (which controls the absolute transit depth) and an opaque, grey, uniformly vertically distributed, cloud opacity ($\kappa_{cld}$). The optimal model resulting from this process is [M/H] = +1.0, C/O = 0.35, and $\log_{10}\kappa_{cld}$ = -2.15 cm$^2$/g. The metallicity and cloud opacity are primarily driven by the strength of the 4.3 μm $CO_2$ feature and lack of $CH_4$ absorption near 3.3 μm. This result is what is shown in the main text (see Figure 3), which also illustrates the relative contribution of the key opacity sources ($H_2O$[75,76], CO[77,78], $CO_2$[79,80], $H_2S$[78,81], and $CH_4$[78,82]) to the overall spectral shape. ED Figure 2 shows the atmospheric structure (temperature profile and gas mixing ratio profiles) for this best fit model.

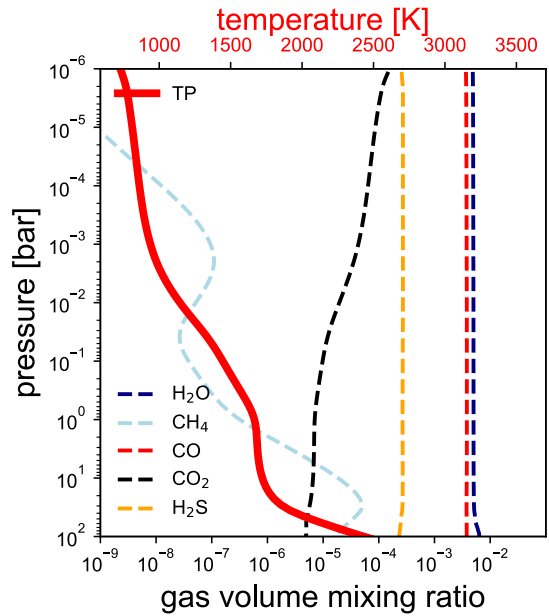

**ED Figure 2: Atmospheric structure arising from the best fit model.** The thick red curve (and corresponding top x-axis) shows the resulting 1D radiative-convective equilibrium temperature profile. The dashed lines (and bottom x-axis) show the vertical gas mixing ratio profiles under the assumption of thermochemical equilibrium. These abundances, along with the absorption cross-sections shown in the bottom panel of Figure 3, are what control the relative contributions of each gaseous opacity to the total transmission spectrum.

### PICASO

The core 1D radiative-convective model is based upon the legacy "Extrasolar Giant Planet (EGP)" code described in Refs. [69, 80, 83] and since updated and modernized within the PICASO[84] framework described in Mukherjee et al., submitted (PICASO 3.0). The PICASO 3.0 model uses gaseous opacities created from the references listed in Ref [80]. The grid of PICASO models contains metallicity points at 0.1, 0.3, 1, 3, 10, 30, 50, and 100x solar; C/O at 0.23, 0.46, 0.69, and 0.92; and also assumes full day-night heat redistribution. The clouds are modeled using the Virga[85] implementation of the Eddysed[86] framework, which requires a vertical mixing coefficient (constant with altitude; $\log_{10} K_{zz}$ = 5, 7, 9, and 11 [cgs units]) and a vertically-constant sedimentation parameter ($f_{sed}$ = 0.6, 1, 3, 6, and 10), with optical/material properties for clouds thought to exist at WASP-39b's pressures and temperatures ($Na_2S$, $MnS$, and $MgSiO_3$). The $f_{sed}$ parameter controls the vertical extent of the cloud, and $K_{zz}$ and $f_{sed}$ together control the mean droplet sizes with altitude in the atmosphere. A chi-square grid search along the described dimensions is performed to identify the best fit. Within this grid, the nominal best fit ($\chi^2/N_{data}$=1.34) is 10x solar metallicity, a sub-solar C/O (0.23), with an extended large droplet cloud ($f_{sed}$ = 0.6, $\log_{10}K_{zz}$ = 9) that produces a grey continuum over these wavelengths, consistent with the ScCHIMERA results above.

### ATMO

The ATMO radiative-convective-thermochemical equilibrium solver is described in Refs [87-90]. This grid consists of model transmission spectra for four different day-night energy redistribution factors (0.25, 0.5, 0.75, 1.0, where 0.5 is "full", and 1.0 is "dayside-only"), six metallicities (0.1, 1, 10, 50, 100, 200 times solar), six C/O ratios (0.35, 0.55, 0.70, 0.75, 1.0, 1.5), two haze factors (no haze and 10 times multi-gas Rayleigh Scattering) and four grey cloud factors (no cloud, 0.5, 1, and 5 times the strength of $H_2$ Rayleigh Scattering at 350 nm between 1 and 50 mbar pressure levels). Each model transmission spectrum from the grid is binned to the same resolution as that of the observations to compute $\chi^2$, with a (wavelength-independent) transit depth offset as the free parameter. Within this grid, we find a best-fit model ($\chi^2/N_{data}$ = 1.39) spectrum arising from a redistribution factor of 0.75 (slightly hotter than a full day-night redistribution would produce), a metallicity of 10x solar, a super-solar C/O ratio of 0.7, a haze factor of 10, and a cloud factor of 5.

### PHOENIX

This model originates from the PHOENIX stellar atmosphere code[91] adapted for exoplanets[92] with additional modeling and opacity updates described in Refs [93,94]. The model grid is computed for an array of irradiation temperatures (920, 1020, 1120, and 1220 K), metallicities (0.1, 1, 10, and 100x solar), C/O (0.3, 0.54, 0.7, and 1.0), and includes a sampling of opaque, grey clouds at specified cloud-top pressures. The nominal best-fit model ($\chi^2/N_{data}$ = 1.32) from this grid setup results in a 10x solar metallicity, subsolar C/O (0.3) atmosphere with a cloud-top pressure of 0.3 mbar.

**Quantifying Feature Detection Significance**

We quantified the detection significance[95] of $CO_2$ with the following steps: 1) the best-fit grid model without $CO_2$ (i.e., the "no $CO_2$" black curve shown in Figure 3) is first subtracted from the data, leaving behind a strong residual feature due to $CO_2$ (ED Figure 3). The peak per-spectral-bin mean SNR of this residual feature is ~10σ. To utilize the full line/band shape we then fit the residual peak with (1) a four-parameter Gaussian model (centroid, amplitude, width, and vertical offset), shown as red curves in ED Figure 3, and (2) a "no feature" constant using a nested sampling routine[74]. The Bayesian evidence between the Gaussian model and constant model were then used to compute a Bayes factor, $B$, and corresponding detection significance[96]. For the $CO_2$ residual feature, $\ln(B)$ is 340.5, which equates to a 26.2σ detection. From this analysis, we conclude that the $CO_2$ feature is robustly detected.

Upon inspecting Figures 2 and 3 in the main text, there appears to be a feature near 4.0 μm (just short of the major $CO_2$ feature). We repeated the same analysis as above, but instead compared the Bayesian evidence from a 2-component Gaussian model fit (to accommodate for both the $CO_2$ feature and the unknown absorber) to that of the single component Gaussian model fit above. Upon doing so, we find ln(*B*) = 0.98 which equates to a 2σ significance. Restricting the prior range for the second Gaussian to be localized near the 4 μm feature boosts the significance to 2.3σ. Future analyses will focus on the nature of this feature and more rigorous quantification via nested Bayesian model comparison within atmospheric retrieval frameworks (e.g., Ref [34]).

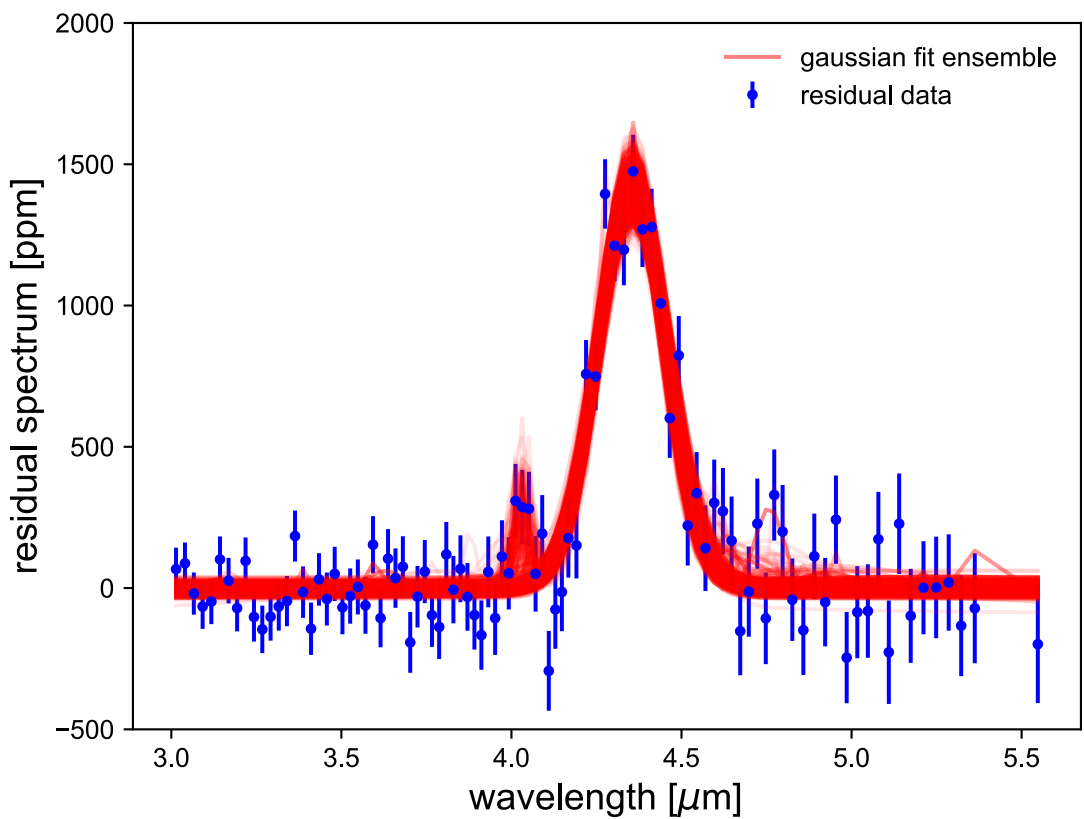

**ED Figure 3: Assessment of the strength of spectral features for WASP-39b.** Residual features (blue data points) after subtracting the continuum best model (black "no $CO_2$" model curve in Figure 3). A best-fitting ensemble of a 2-component Gaussian model to both the $CO_2$ feature and the unknown absorber feature (~4 μm) is shown in red.

**Data Availability**

The data used in this paper are associated with JWST program ERS 1366 (observation #4) and are available from the Mikulski Archive for Space Telescopes (https://mast.stsci.edu). Science data processing version (SDP_VER) 2022_2a generated the uncalibrated data that we downloaded from MAST. We used JWST calibration software version (CAL_VER) 1.5.3 with modifications described in the text. We used calibration reference data from context (CRDS_CTX) 0916, except as noted in the text. All the data and models presented in this publication can be found at https://doi.10.5281/zenodo.6959427.

**Code Availability**

The codes used in this publication to extract, reduce and analyse the data are as follows; STScI JWST Calibration pipeline[37] (https://github.com/spacetelescope/jwst), tshirt[40], Eureka![39] (https://eurekadocs.readthedocs.io/en/latest/), Tiberius[21,56,57], FIREFLy[38]. In addition, these made use of Exoplanet[51] (https://docs.exoplanet.codes/en/latest/), Pymc3[52] (https://docs.pymc.io/en/v3/index.html), ExoTEP[53-55], Batman[56] (http://lkreidberg.github.io/batman/docs/html/index.html), ExoTiC-ISM[59] (https://github.com/Exo-TiC/ExoTiC-ISM), ExoTiC-LD[60] (https://exotic-ld.readthedocs.io/en/latest/), Emcee[63] (https://emcee.readthedocs.io/en/stable/), Dynesty[74] (https://dynesty.readthedocs.io/en/stable/index.html), and chromatic (https://zkbt.github.io/chromatic/), each of which use the standard python libraries scipy[97], numpy[98], astropy[99,100], and matplotlib[101]. The atmospheric models used to fit the data can be found at PICASO[84] (https://natashabatalha.github.io/picaso/), Virga[85] (https://natashabatalha.github.io/virga/), ScCHIMERA[68] (https://github.com/mrline/CHIMERA), ATMO[87-90], and PHOENIX[93].

**Methods References**

**Acknowledgements** This work is based on observations made with the NASA/ESA/CSA James Webb Space Telescope. The data were obtained from the Mikulski Archive for Space Telescopes at the Space Telescope Science Institute, which is operated by the Association of Universities for Research in Astronomy, Inc., under NASA contract NAS 5-03127 for JWST. These observations are associated with program JWST-ERS-01366. Support for program JWST-ERS-01366 was provided by NASA through a grant from the Space Telescope Science Institute, which is operated by the Association of Universities for Research in Astronomy, Inc., under NASA contract NAS 5-03127. The results reported herein benefited during the design phase from collaborations and/or information exchange within NASA's Nexus for Exoplanet System Science (NExSS) research coordination network sponsored by NASA's Science Mission Directorate.

**Author contributions** All members of the JWST Transiting Exoplanet Community Early Release Science Team played a significant role in one or more of the following: development of the original proposal, management of the project, definition of the target list and observation plan, analysis of the data, theoretical modeling, and preparation of this manuscript. Some specific contributions are listed as follows. NMB, JLB, and KBS provided overall program leadership and management. DS, EK, HW, IC, JLB, KBS, LK, MLM, MRL, NMB, VP, and ZBT made significant contributions to the design of the program. KBS generated the observing plan with input from the team. EvSc, NE, and TGB provided instrument expertise. BB, EK, HW, IC, JLB, LK, MLM, MRL, NMB, and ZBT led or co-led working groups and/or contributed to significant strategic planning efforts like the design and implementation of the pre-launch Data Challenges. AC, DS, EvSc, NE, NG, TGP, VP generated simulated data for pre-launch testing of methods. AC, AF, CP, EA, EvSc, JaK, LA, TJB, and ZR contributed to the development of data analysis pipelines and/or provided the data analysis products used in this analysis. Reduced the data, modeled the light curves, and produced the planetary spectrum. JG, JoL, KO, MRL, NEB, SaMu, and SEM generated theoretical model grids for comparison with data. AC, DS, EvSc, HW, JaK, JF, JG, JLB, JoL, KBS, KO, MRL, NEB, NMB, SaMu, SEM, ZBT, and ZR contributed significantly to the writing of this manuscript. EA, EvSc, MRL, and ZBT generated figures for this manuscript.

**Competing interests** The authors declare no competing interests.

**Additional information** none
**Supplementary information** none

**Correspondence and requests for materials** should be addressed to Natalie Batalha (Natalie.Batalha@ucsc.edu).

The JWST Transiting Exoplanet Community Early Release Science Team:

Eva-Maria Ahrer[1,2], Lili Alderson[3], Natalie M. Batalha[4], Natasha E. Batalha[5], Jacob L. Bean[6], Thomas G. Beatty[7], Taylor J. Bell[8], Björn Benneke[9], Zachory K. Berta-Thompson[10], Aarynn L. Carter[4], Ian J. M. Crossfield[11], Néstor Espinoza[12,13], Adina D. Feinstein[6,14], Jonathan J. Fortney[4], Neale P. Gibson[15], Jayesh M. Goyal[16], Eliza M.-R. Kempton[17], James Kirk[18], Laura Kreidberg[19], Mercedes López-Morales[18], Michael R. Line[20], Joshua D. Lothringer[21], Sarah E. Moran[22], Sagnick Mukherjee[4], Kazumasa Ohno[4], Vivien Parmentier[23,24], Caroline Piaulet[9], Zafar Rustamkulov[25], Everett Schlawin[26], David K. Sing[25,13], Kevin B. Stevenson[26], Hannah R. Wakeford[3], Natalie H. Allen[13,14], Stephan M. Birkmann[27], Jonathan Brande[11], Nicolas Crouzet[28], Patricio E. Cubillos[29,30], Mario Damiano[31], Jean-Michel Désert[32], Peter Gao[33], Joseph Harrington[34], Renyu Hu[31,35], Sarah Kendrew[27], Heather A. Knutson[35], Pierre-Olivier Lagage[36], Jérémy Leconte[37], Monika Lendl[38], Ryan J. MacDonald[39], E. M. May[26], Yamila Miguel[28,40], Karan Molaverdikhani[41,42,19], Julianne I. Moses[43], Catriona Anne Murray[10], Molly Nehring[10], Nikolay K. Nikolov[12], D. J. M. Petit dit de la Roche[38], Michael Radica[9], Pierre-Alexis Roy[9], Keivan G. Stassun[44], Jake Taylor[9], William C. Waalkes[10], Patcharapol Wachiraphan[10], Luis Welbanks[20,45], Peter J. Wheatley[2,1], Keshav Aggarwal[46], Munazza K. Alam[33], Agnibha Banerjee[47], Joanna K. Barstow[47], Jasmina Blecic[48], S.L. Casewell[49], Quentin Changeat[50], K. L. Chubb[51], Knicole D. Colón[52], Louis-Philippe Coulombe[9], Tansu Daylan[53,54], Miguel de Val-Borro[55], Leen Decin[56], Leonardo A. Dos Santos[12], Laura Flagg[39], Kevin France[57], Guangwei

Fu[17,13], A. García Muñoz[36], John E. Gizis[58], Ana Glidden[59,60], David Grant[3], Kevin Heng[61], Thomas Henning[19], Yu-Cian Hong[39], Julie Inglis[35], Nicolas Iro[62], Tiffany Kataria[31], Thaddeus D. Komacek[17], Jessica E. Krick[63], Elspeth K.H. Lee[64], Nikole K. Lewis[39], Jorge Lillo-Box[65], Jacob Lustig-Yaeger[26], Luigi Mancini[66,19,67], Avi M. Mandell[52], Megan Mansfield[26,45], Mark S. Marley[22], Thomas Mikal-Evans[19], Giuseppe Morello[68,69,70], Matthew C. Nixon[71], Kevin Ortiz Ceballos[18], Anjali A. A. Piette[33], Diana Powell[18], Benjamin V. Rackham[59,60,72], Lakeisha Ramos-Rosado[13], Emily Rauscher[73], Seth Redfield[74], Laura K. Rogers[71], Michael T. Roman[49,75], Gael M. Roudier[31], Nicholas Scarsdale[4], Evgenya L. Shkolnik[20], John Southworth[76], Jessica J. Spake[35], Maria E Steinrueck[19], Xianyu Tan[23], Johanna K. Teske[33], Pascal Tremblin[77], Shang-Min Tsai[23], Gregory S. Tucker[78], Jake D. Turner[39,46], Jeff A. Valenti[12], Olivia Venot[79], Ingo P. Waldmann[50], Nicole L. Wallack[35], Xi Zhang[80], Sebastian Zieba[19,28]

[1]Department of Physics, University of Warwick, Coventry, UK
[2]Centre for Exoplanets and Habitability, University of Warwick, Coventry, UK
[3]School of Physics, University of Bristol, Bristol, UK
[4]Department of Astronomy and Astrophysics, University of California, Santa Cruz, Santa Cruz, CA, USA
[5]NASA Ames Research Center, Moffett Field, CA, USA
[6]Department of Astronomy & Astrophysics, University of Chicago, Chicago, IL, USA
[7]Department of Astronomy, University of Wisconsin-Madison, Madison, WI USA
[8]BAER Institute, NASA Ames Research Center, Moffet Field, CA, USA
[9]Department of Physics and Institute for Research on Exoplanets, Université de Montréal, Montreal, QC, Canada
[10]Department of Astrophysical and Planetary Sciences, University of Colorado, Boulder, CO, USA
[11]Department of Physics & Astronomy, University of Kansas, Lawrence, KS, USA
[12]Space Telescope Science Institute, Baltimore, MD, USA
[13]Department of Physics & Astronomy, Johns Hopkins University, Baltimore, MD, USA
[14]NSF Graduate Research Fellow
[15]School of Physics, Trinity College Dublin, Dublin, Ireland
[16]School of Earth and Planetary Sciences (SEPS), National Institute of Science Education and Research (NISER), HBNI, Odisha, India
[17]Department of Astronomy, University of Maryland, College Park, MD, USA
[18]Center for Astrophysics | Harvard & Smithsonian, Cambridge, MA, USA
[19]Max Planck Institute for Astronomy, Heidelberg, Germany
[20]School of Earth and Space Exploration, Arizona State University, Tempe, AZ, USA
[21]Department of Physics, Utah Valley University, Orem, UT, USA
[22]Lunar and Planetary Laboratory, University of Arizona, Tucson, AZ, USA.
[23]Atmospheric, Oceanic and Planetary Physics, Department of Physics, University of Oxford, Oxford, UK
[24]Université de Nice-Sophia Antipolis, Observatoire de la Côte d'Azur, Nice, France

[25]Department of Earth and Planetary Sciences, Johns Hopkins University, Baltimore, MD, USA
[26]Steward Observatory, University of Arizona, Tucson, AZ, USA
[27]Johns Hopkins APL, Laurel, MD, USA
[28]European Space Agency, Space Telescope Science Institute, Baltimore, MD, USA
[29]Leiden Observatory, University of Leiden, Leiden, The Netherlands
[30]INAF – Osservatorio Astrofisico di Torino, Pino Torinese, Italy
[31]Space Research Institute, Austrian Academy of Sciences, Graz, Austria
[32]Astrophysics Section, Jet Propulsion Laboratory, California Institute of Technology, Pasadena, CA, USA
[33]Anton Pannekoek Institute for Astronomy, University of Amsterdam, Amsterdam, The Netherlands
[34]Earth and Planets Laboratory, Carnegie Institution for Science, Washington, DC, USA
[35]Planetary Sciences Group, Department of Physics and Florida Space Institute, University of Central Florida, Orlando, Florida, USA
[36]Division of Geological and Planetary Sciences, California Institute of Technology, Pasadena, CA, USA
[37]Université Paris-Saclay, Université Paris Cité, Gif-sur-Yvette, France
[38]Laboratoire d'Astrophysique de Bordeaux, Université de Bordeaux, Pessac, France
[39]Département d'Astronomie, Université de Genève, Sauverny, Switzerland
[40]Department of Astronomy and Carl Sagan Institute, Cornell University, Ithaca, NY, USA
[41]SRON Netherlands Institute for Space Research, Leiden, the Netherlands
[42]Universitäts-Sternwarte, Ludwig-Maximilians-Universität München, München, Germany
[43]Exzellenzcluster Origins, Garching, Germany
[44]Space Science Institute, Boulder, CO, USA
[45]Department of Physics and Astronomy, Vanderbilt University, Nashville, TN, USA
[46]NHFP Sagan Fellow
[47]Indian Institute of Technology, Indore, India
[48]School of Physical Sciences, The Open University, Milton Keynes, UK
[49]Department of Physics, New York University Abu Dhabi, PO Box 129188 Abu Dhabi, UAE
[50]School of Physics and Astronomy, University of Leicester, Leicester
[51]Department of Physics and Astronomy, University College London, United Kingdom
[52]Centre for Exoplanet Science, University of St Andrews, St Andrews, UK
[53]NASA Goddard Space Flight Center, Greenbelt, MD, USA
[54]Department of Astrophysical Sciences, Princeton University, Princeton, NJ, USA
[55]LSSTC Catalyst Fellow
[56]Planetary Science Institute, Tucson, AZ, USA
[57]Institute of Astronomy, Department of Physics and Astronomy, KU Leuven, Leuven, Belgium
[58]Laboratory for Atmospheric and Space Physics, University of Colorado Boulder, Boulder, CO, USA
[59]Department of Physics and Astronomy, University of Delaware, Newark, DE, USA

[60]Department of Earth, Atmospheric and Planetary Sciences, Massachusetts Institute of Technology, Cambridge, MA, USA

[61]Kavli Institute for Astrophysics and Space Research, Massachusetts Institute of Technology, Cambridge, MA, USA

[62]University Observatory Munich, Ludwig Maximilian University, Munich, Germany

[63]Institute for Astrophysics, University of Vienna, Vienna, Austria

[64]Astrophysics Section, Jet Propulsion Laboratory, California Institute of Technology, Pasadena, CA, USA

[65]Center for Space and Habitability, University of Bern, Bern, Switzerland

[66]Centro de Astrobiología (CAB, CSIC-INTA), Departamento de Astrofísica, ESAC campus, Villanueva de la Cañada (Madrid), Spain

[67]Department of Physics, University of Rome "Tor Vergata", Rome, Italy

[68]INAF - Turin Astrophysical Observatory, Pino Torinese, Italy

[69]Instituto de Astrofísica de Canarias (IAC), Tenerife, Spain

[70]Departamento de Astrofísica, Universidad de La Laguna (ULL), Tenerife, Spain

[71]INAF- Palermo Astronomical Observatory, Piazza del Parlamento, Palermo, Italy

[72]Institute of Astronomy, University of Cambridge, Cambridge, UK

[73]51 Pegasi b Fellow

[74]Department of Astronomy, University of Michigan, Ann Arbor, MI, USA

[75]Astronomy Department and Van Vleck Observatory, Wesleyan University, Middletown, CT, USA

[76]Universidad Adolfo Ibáñez: Penalolen, Santiago, CL

[77]Astrophysics Group, Keele University, Staffordshire, UK

[78]Maison de la Simulation, CEA, CNRS, Univ. Paris-Sud, UVSQ, Université Paris-Saclay, Gif-sur-Yvette, France

[79]Department of Physics, Brown University, Providence, RI, USA

[80]Université de Paris Cité and Univ Paris Est Creteil, CNRS, LISA, Paris, France

[81]Department of Earth and Planetary Sciences, University of California Santa Cruz, Santa Cruz, California, USA